# Strong coupling and high contrast all optical modulation in atomic cladding waveguides

**Liron Stern,[1] Boris Desiatov,[1] Noa Mazurski,[1] and Uriel Levy[1,*]**

[1]Department of Applied Physics, The Benin School of Engineering and Computer Science, The Center for Nanoscience and Nanotechnology, The Hebrew University of Jerusalem, Jerusalem, 91904, Israel
*Corresponding author: ulevy@cc.huji.ac.il

**In recent years we are witnessing a flourish in research aimed to facilitate alkali vapors in guided wave configurations. Owing to the significant reduction in device dimensions, the increase in density of states, the interaction with surfaces and primarily the high intensities carried along the structure, a rich world of light vapor interactions can be studied, and new functionalities, e.g. low power nonlinear light-matter interactions can be achieved. One immense remaining challenge is to study the effects of quantum coherence and shifts in such nano-scale waveguides, characterized by ultra-small mode areas and fast dynamics. Here, we construct a serpentine silicon-nitride wave guide, having atomic vapor as its cladding. The unprecedented mode volume of $5 \cdot 10^{-13}$ m$^3$ supported over a length of 17 mm is used to demonstrate efficient linear and non-linear spectroscopy. Fascinating and important phenomena such as van der Waals shifts, dynamical stark shifts, and coherent effects such as strong coupling (in the form of Autler Townes splitting) are all observed. The serpentine atomic cladding is a promising building block for a variety of light vapor experiments, as it offers a very small footprint, enables operation with relatively low density of atoms and extremely strong confinement of light and vapor. As such it may be used for important applications, such as all optical switching, frequency referencing, and magnetometry to name a few.**

The emergence of the field of guided light vapor interactions, has led to numerous demonstrations of efficient linear and nonlinear light vapor interactions[1–11]. Motivated by the general trend for miniaturization and high level integration, this effort strives to confine light and vapor in the same quarters in order to obtain efficient and low-power light-vapor interactions. Integrating vapor cells with optical fibers and waveguides is natural choice for enhancing light vapor interactions. This is because guided waves do not exhibit light diffraction as an optical waveguide supports tight mode confinement over a long propagation distance. A few approaches for integrating vapors with guided wave configurations have been demonstrated over the recent years, including the use of hollow core photonic crystal fibers[3,4,9,10] (HC-PCF), hollow core anti reflecting optical wave-guides[12,13] (HC-ARROW), tapered nano-fibers[1,2,11] (TNF), surface-plasmon guided resonances[7] and exposed core fibers[5]. Such integration of vapor with guided systems has served for the demonstration of a myriad of effects ranging from basic linear spectroscopy[1,3,6,12], to a variety of non-linear effects such as electromagnetic induced transparency[14] (EIT), enhanced two-photon absorption[2,10], phase switching[4], all optical modulation[7,15] and slow light[13].

Recently we have demonstrated the atomic cladding wave guide (ACWG), consisting of a silicon nitride (SiN) core surrounded by a cladding of Rb vapor which is introduced by integrating an atomic vapor cell above the optical chip[6]. In this demonstration, the ACWG had an interaction length of 1.5mm and a mode area of $0.3\lambda^2$. With this configuration we demonstrated basic spectroscopy, extremely low saturation power, and two photon transitions. Subsequently, the integration of ACWG based micro ring resonators has been demonstrated to achieve efficient all optical modulation[15]. Even more recently, Ritter et. al. have implemented such ACWGs in a Mach Zhender interferometer configuration[16]. Such a platform offers several prominent advantages; dimension-wise, an ACWG, has a mode area that is about two orders of magnitude smaller than previously demonstrated on chip approaches, making it a fine candidate for efficient non-linear light vapor interactions. In addition, due to the ability to integrate large variety of existing photonic circuits with the ACWG, this approach can further enhance both linear and non-linear light-matter interaction via the use of resonators such as micro-ring resonators and photonic crystal resonators, and structures such as slot wave guides. The ACWG silicon nitride platform further enables to couple broadband optical signals, making it applicable (simultaneously) to various alkali vapors such as Cs and Rb and various optical transitions whether the D1 and D2 lines, or higher excited Rydberg states[17].

Here, we introduce the platform of a long serpentine-like atomic cladding wave guide for the purpose of demonstrating two major phenomena: **The first**, is the ability to obtain coherent effects, using low optical powers for the purpose of all optical switching. Indeed, we have been able to almost "switch off" a probe signal, in the presence of a pump beam, having power of only 10 μW. Such switching capability is expected to be relatively fast, as the dynamics of our hot vapor apparatus are in the nano-second regime. **The second**, is the ability to quantify two important mechanisms for shifts, namely the Van der Waals (VDW) shift and the light shift. Understanding such effects is crucially important for the construction of on-chip frequency references using such platforms.

Specifically, owing to our design, which offers high optical density on chip with an ultra-small footprint we demonstrate linear absorption approaching 25% at temperature as low as 65°C. Due to the high electromagnetic energy density propagating in the waveguide we demonstrate that such a platform retains its previous merits, and saturates at the 100nW regime. Moreover, when introducing a pump field, we are able to observe **strong coupling** in the form of a high contrast Autler-Townes splitting, which can be exploited for highly efficient all optical switching, using microwatts of power levels. In addition, at such high intensity levels, we observe light shifts of approximately 200MHz. Compared to previous demonstrations, the current device achieves enhanced performance in terms of optical density, and non-linearities. As such, it holds an immense promise as a fundamental building block in a variety of light-vapor experiments. Finally, our device offers extremely fast (sub ns) transit times, as a result of the exceptionally small evanescent decay length of the optical mode away from the waveguide core. Such fast transit times accompanied with fast induced Rabi frequencies allow one to observe coherent effects, and ultrafast switching speeds, which are highly advantageous in applications such as all optical switching.

A sketch of our atomic cladding waveguide is depicted in figure 1a, whereas a schematic cross section is shown in Fig 1b. The ACWG is 17mm long, and sufficient waveguide spacing of ~10 microns is used to avoid coupling of light between adjacent pieces of the same waveguide. The waveguide is constructed by standard electron beam lithography followed by reactive ion etching. Its dimensions were designed to support a relative broad spectral range, both in the visible and in the telecom regime. Thus the waveguide supports the D1 and D2 transitions of Rb as well as non-direct transitions at 1319nm and 1529nm. Following the definition of the waveguide core we deposit 1μm thick layer of SiO2 on the entire chip using plasma enhanced chemical vapor deposition (PECVD). Next we use coarse lithography followed by wet etching to expose the waveguides top cladding. Finally, we integrate the chip with a vapor cell, similarly to the procedure reported in ref. 6. In short, we bond a flat hollow cylinder to the chip with thermal cured vacuum grade epoxy, bake-out and evacuate the bonded cell, introduce $^{85}$Rb into the cell and pinch-off the glass cell from the vacuum system to obtain a portable device.

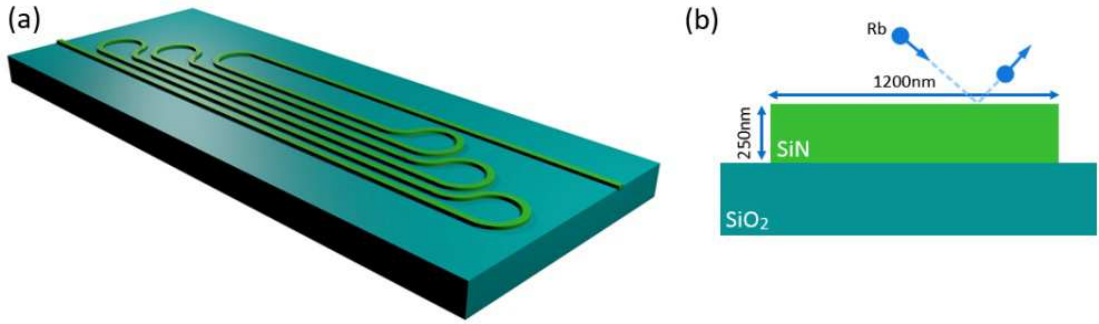

**Fig. 1.** (a) Sketch of a serpentine atomic cladding wave guide consisting of seven segments of swirled SiN waveguides (b) Cross section sketch of the silicon nitride wave guide, of 250nm height and 1200nm width. Rb atoms are illustrated as balisiticaly striking the surface, entering and exiting the evanescent portion of the optical mode.

Next we couple light directly to the ACWG to perform spectroscopic measurements. To do so, we use lensed fibers for coupling light from a fiber coupled 780nm laser into the waveguide, and out of the waveguide to the detector. We use the D2 line of rubidium, around 780nm wavelength. The normalized transmission spectrum of light propagating along the ACWG is shown in Fig 2a, together with a transmission spectrum of a reference cell. The results were obtained at cell's temperature of about $65^oC$, corresponding to atomic density of $3.7 \cdot 10^{11}$ $cm^{-3}$. The optical power in the waveguide was attenuated in order to avoid full saturation of the transitions, and is estimated to be about 50nW. As can be clearly seen, in comparison with the reference cell the absorption lines are broader. As reported[6,16,18], this result is associated with an additional Doppler broadening due to the increased photon momentum along the propagation direction, and also with an increase in the transit time broadening resulted by the limited interaction time of the vapor with the evanescent tail of the optical mode. This broadening manifests as a unified line (consisting of three different unresolved transitions) with a full width half maximum of ~1.3 GHz. The mode content within the waveguide directly affects the Doppler broadening via the effective refractive index. While the exact distribution of modes is unknown, we have assumed that the electromagnetic energy is equally distributed among the first 3 modes, and thus the average effective index is ~1.66 (calculated by finite element method simulation), with an uncertainty lower than 5%. As a result, the Doppler broadened linewidth is estimated to be ~1.05GHz. The remaining broadening of 250 MHz is mostly attributed to transit time broadening.

Next we compare the spectral position of the dips in respect to the reference cell. Interestingly, we observe a red-shift of approximately 65 MHz. This redshift may originate from long range van der Waals (VDW) interaction between the Rb atoms and the silicon nitride surface. As the average evanescent length of the first three modes is approximately 90nm, using an effective interaction length of 45nm[19], yields a VDW coefficient of 6KHz/$\mu m^{-3}$, being in the range of VDW coefficients reported previously[20,21]. Exploring surface interactions such as VDW shifts is important both from fundamental and applicative aspects. For instance, in metrology applications such as optical frequency references, exploring the VDW coefficient for different types of materials, and coatings as well as its environmental (e.g. temperature) dependency is highly important. Further work will be devoted to this topic in the future.

Taking into account the above mentioned effects, we can now fit the obtained spectrum to a model. This model includes transit time broadening, enhanced Doppler broadening, the quenching of atoms on the waveguides wall, the slight saturation of the atoms and the VDW shift. More details about this model are given in ref. 22 and in 6. In this model, a single fitting parameter is used - the density of atoms (or equivalently, the temperature). The fitting results are displayed in Fig 2a (see red curve). It is important to note that owing to the use of a relatively long waveguide a high contrast of ~20% is now achievable at a moderate temperature of 65$^\circ$, which is a significant improvement compared with previously reported results achieved with shorter waveguides. Furthermore, the compact serpentine design allows to keep the footprint of the device as small as possible.

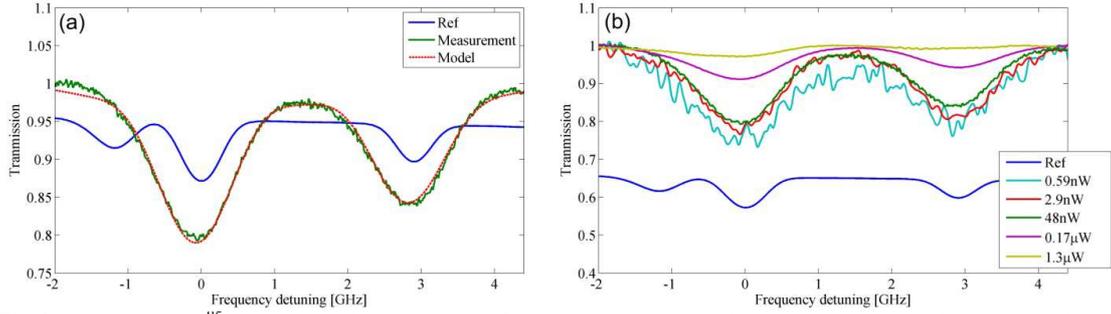

**Fig 2**: (a) Measured $^{85}$Rb transmission spectrum of a 17 mm long ACWG (green), together with the calculated transmission spectrum (red) and the measured transmission spectrum of reference cell containing natural Rb (blue). (b) Measured ACWG spectra at different power levels within the waveguide, together with the measured transmission spectrum of reference cell containing natural Rb (blue)

Next we vary the coupled power in the ACWG and measure its transmission spectrum. Five different measured spectra corresponding to five different power levels are plotted in Fig 2b. Clearly, it can be seen that varying the power levels changes the transmission spectrum. Specifically, the transition contrasts are diminished for powers of few µWs while for power levels in the nW regime these transitions are recovered. This is a direct consequence of the two-level saturation mechanism. The onset of saturation is estimated to be in the 100nW regime, consistent with previously reported results[1,2,6]. This low saturation power is due to the high energy density carried by the guided mode, and serves as an indication for the low light level needed for other non-linear process such as two photon absorption, and EIT.

One significant manifestation of such non-linear process is all optical switching, i.e. the ability to modulate light transmission at a particular wavelength in the presence of signal at another wavelength by a non-linear mediator. Amongst many different possible schemes for all optical switching in Rb, we implement here all optical switching in the D1 and D2 wavelengths. A schematic representation of this process is shown in Fig 3a, whereas a schematic representation of the optical configuration is presented in Fig. 3b. First, we tune the D2 line of 780nm to the F=3 to F'=2/3/4 transition, with a pump power in the µW regime within the waveguide. Next, we scan the D1 line (with a relatively unsaturated power level of the order of 100nW) of 795nm across the F=2/3 to F'=3/2 manifolds. Both wavelengths co-propagate, and are coupled into the waveguide in the same manner described earlier. The out coupled light, collected with a lensed fiber, is collimated and spectrally filtered using two band pass filters, operating at the ~800nm band, and thus filtering out the 780nm wavelength. Finally, as before, we collect the filtered signal using a photo detector.

Before elaborating on the results obtained in the waveguide, we briefly describe the results obtained simultaneously using a reference cell (Figure 3c). Obviously, for this cm-size cell operating with free space beams the intensity levels for both probe and pump are much lower with respect to the waveguide. In Fig. 3c we plot the spectrum of the probe beam with (green line) and without (blue line) the presence of the 780 pump beam. Clearly, we can see that pump beam changes the transmission spectrum drastically. First, we observe peaks within the F=3 to F'=2/3 transition, attributed to combination of V-type EIT, velocity selective optical pumping and saturation[23,24]. Next, we observe an enhancement of the absorption for the F=2 to F'=2/3 transition, in the form of two dips[7]. We attribute this absorption enchantment to optical pumping which is induced by the pump beam on the F=3 to F'=2/3/4 manifold. We note that these peaks and dips are power broadened, and it should be possible to observe nearly natural line widths (~6MHz) using this scheme by reducing the pump power, albeit at the expanse of a lower contrast.

Following this discussion, we can now turn back to describe the atomic cladding waveguide pump-probe spectra. First, we plot in Fig. 3d the transmission spectrum of the probe beam with (green line) and without (blue line) the presence of the 780 pump beam. Here, we use a pump beam with power of approximately 10 µW within the wave guide, and an unsaturated probe beam. As can be seen the pump beam strongly modulates the probe spectrum. When the probe is tuned to the F=3 to F'=2/3 line (around zero detuning), we observe a drastic increase in transmission, whereas when the probe is tuned to the F=2 to F'=2/3 line (around 3 GHz detuning), we do not observe any significant change in transmission, and yet observe a significant frequency shift of approximately 200 MHz. We note that for this pump power we are able to almost totally "shut-down" the absorption of the probe, accompanied with an appearance of a transparency window within the absorption dip. We shall claim that this is a direct consequence of atomic coherence induced by the pump beam. In the case of the F=3 to F'=2/3 transition, the pump depletes the occupancy of F=3 level, and as a result there is less absorption and increased transmission. In contrast, for the F=2 to F'=2/3 transition the atoms which are pumped from the F=3 level to the F''=2/3 manifold can spontaneously decay to the F=2 level, and thus increase the population of level F=2 and increase absorption. The magnitude of this phenomena is relatively small due to: a) The lifetime of 27ns of Rb is larger than the average time the atom spends in the evanescent region b) the branching ratio of decays routs which is favorable towards the F=3 level. The frequency shift, that is observed in the F=2 to F'=2/3 transition is attributed to the dynamical (AC) stark shift, that the pump beam is applying to the F=2 ground state, detuned approximately 3 GHz from the pump frequency.

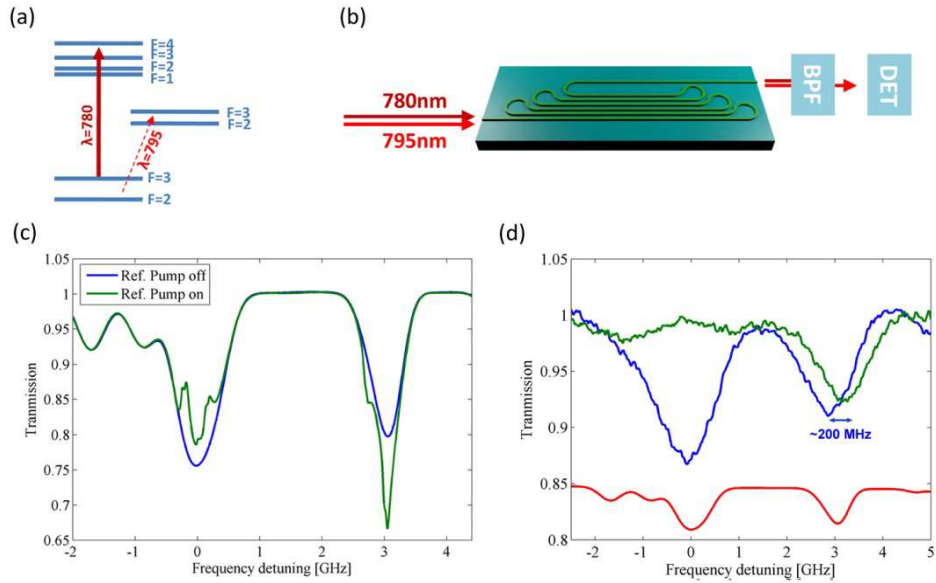

**Fig. 3:** a) $^{85}$Rb relevant transitions of the D1 and D2 manifolds b) Sketch of the optical configuration, illustrating the co-propagating 780nm and 795nm light beams coupled into a serpentine ACWG c) Measured transmission D1 spectrum (795 nm), propagating through a natural Rb cell reference with and without the presence of a pump beam at 780nm d) Measured D1 transmission spectrum, of light propagating through the serpentine ACWG (filled with $^{85}$Rb) with and without the presence of a pump beam at 780nm. The red line represents a D1 lines reference spectrum.

Next, the results obtained for 5 different power levels ranging from 0 to ~13 µW are presented in Fig. 4a:

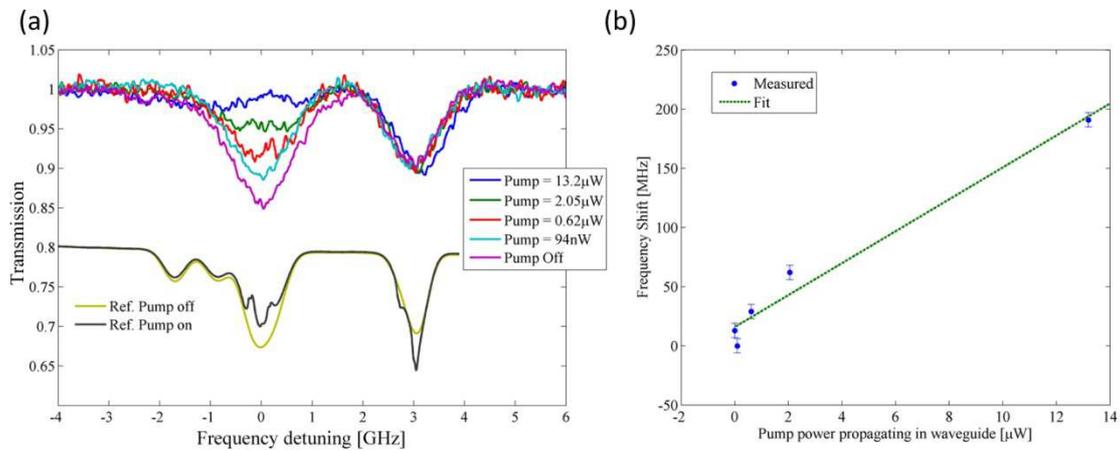

**Fig. 4:** a) Measured transmission spectrum of light around the D1 (795 nm) line, propagating through the serpentine ACWG with varying powers in the presence of a pump beam at 780nm. b) Light shift: F=2 to F'=2/3 dips frequency shift as function of pump power.

As can be observed from Fig. 4a, the contrast of the absorption signal decreases as we increase the power launched in the waveguide, and for the highest power level (blue line, in Fig. 4a), we are able to almost "turn off" the signal. In the absence of optical pumping and coherent effects the maximal achievable population transfer of the pump can be 50%, and thus the almost 100% absorption reduction seems surprising, as it indicates that either optical pumping or coherent effects (or both) must be present. Yet, optical pumping is of negligible magnitude in our system, mostly due to the short transit time. Indeed, we do not observe any significant optical pumping when the probe beam is tuned from the F=2 to F'=2/3 levels. This is in contrast to the optical pumping observed in the reference cell manifested as enhanced absorption (Fig. 4a black line). Thus we claim, and validate numerically, that the significant reduction in the absorption is attributed to a coherent effect. This reduction in contrast is also accompanied with an appearance of broad transparency window. These finding are validated by calculating (see supplementary 1) the susceptibility of the vapor in the case of a three level system and in the frame work of the spatial dependent Bloch equations in the presence of evanescent electromagnetic fields. Indeed, when one discerns numerically between the coherent (represented by the appropriate off diagonal matrix element) and non-coherent process it turns out that our results can only be explained by the presence of such coherent effects. As our system is governed by fast dynamics, due to the short transit time of the atom in the evanescent region, the observation of such coherent process requires strong intensities, corresponding to Rabi frequencies comparable to the transit time frequency (~GHz). Indeed, owing to the highly confined waveguide mode we can achieve Rabi frequencies as high as ~1 GHz along the entire 17 mm of propagation, with moderate optical power of ~10 µW within the waveguide.

As mentioned before, the enhanced transparency window is accompanied by a broad peak. This peak is approximately 1 GHz wide, and is attributed to a strong coupling effect, i.e. the Autler-Townes splitting. Indeed, such an effect is observed for cases of very high Rabi frequency $\Omega_r > \Gamma_d$ where $\Gamma_d$ is the Doppler width. We have observed such peaks very consistently, with varying magnitudes, as a function of the specific pump power within the waveguide. We note that it may be possible to achieve narrower peaks, in the electromagnetic induced transparency regime, as has been demonstrated in HC-PCFs[25], simply by reducing the power.

Finally, to quantify the above mentioned light shift, we plot in Fig, 4b the F=2 to F'=2/3 frequency shift (obtained by fitting a Gaussian to the measured transmission dip and identifying its center) as a function of the estimated power which is coupled into the waveguide. The standard expression for the energy shift of an atomic level, induced by an optical field can be calculated using second order perturbation theory. For the case of low intensity relative to the detuning $\Delta$ i.e. $\Omega_r^2/\Delta^2 \ll 1$, a linear[26] dependency between the frequency shift and the optical power is expected. As the detuning in our case is 3 GHz, and the Rabi frequencies (Corresponding to the intensities propagating within the waveguide) vary from the MHz regime to the GHz regime, we expect to find only a slight deviation from the linear regime. By fitting the results to a linear curve we obtain a slope of 13.5 MHz/µW. Using the linear relation between the squared Rabi frequency and the frequency shift (namely $\Delta f = 1/4 \cdot \Omega_r^2/\Delta$) we infer a Rabi frequency of 420 MHz per (µW)$^{1/2}$, very close to the estimated theoretical F=2 to F'=3 Rabi frequency of 405 MHz/(µW)$^{1/2}$ (see supplementary 2). Moreover, at power of ~13.2µW this Rabi frequency of 1.52 GHz is extremely close to the splitting of 1.59 GHz extracted from Fig. 4a. We note that that in case of a Doppler broadened medium, the splitting is expected to be somewhat larger than the Rabi Frequency as evident from ref. 27 and our simulations in supplementary 1. As the splitting is governed by a Rabi frequency corresponding to the F=3 to F' manifold, which in average is larger from the F=2 to F' manifold, this agreement between the splitting separation and the inferred light-shift Rabi frequency is justified. Finally, we stress that a huge frequency shift of ~200 MHz is observed for a moderate power level of only ~13 µW, and with a relatively large detuning of 3 GHz.

To Summarize, we have demonstrated a platform for chip scale and efficient light vapor interactions, consisting of compact and long SiN waveguide interacting evanescently with Rb atoms, and used it to study two major phenomena, namely coherent effects (e.g. strong coupling) and light and surface dependent frequency shifts.

The coherent effects were observed by the strong modulation of a probe light in response to a pump. With the enhancement of the pump power, we could even reach the regime of strong coupling, manifested as an Autler Townes splitting. This phenomenon is a direct consequence of the light induced coherency in our system. Taking advantage of the tight mode confinement over the entire waveguide length allows us to achieve high Rabi frequencies and all optical switching in the µW regime. The speed of such optical switch is in general limited by the life time of the F"=2/3/4 levels (i.e. the $5^2P_{3/2}$ manifold). However, this natural life time of 27ns is larger than the time an average atom traverses the evanescent region of the waveguide, being in the ns regime. Thus, effectively, and as has been previously explained by Salit et al.[28], this would be the time constant for switching corresponding to the hundreds of MHz operation.

Two fascinating mechanisms for frequency shifts were studied. First, we have observed VDW shift of up to 60 MHz. This observation is attributed to the strongly decaying evanescent field outside of the waveguide core. As a result, the field probes atoms which are in close proximity to the silicon nitride surface, experiencing notable long range VDW shifts. We estimate the VDW coefficient to be $6KHz/\mu m^{-3}$. Furthermore, by introducing the pump beam we were able to observe light shifts as high as 200 MHz corresponding to a huge slope of ~13.5 MHz/µW. Studying frequency shifts in such platforms is detrimental for the implementation of metrology applications on a chip. Further investigation of the magnitude of such effects in other materials and environmental conditions (e.g. temperature) is important for example for understanding the ultimate frequency stability of an optical frequency reference in an ACWG. In particular, it would be interesting to study the VDW shift in micro and nano structured materials, where phonons can be controlled.

Before concluding, it is now time to discuss some of the future prospects in the field. In order to fully utilize the flexibility of the silicon photonics platform for the purpose of an efficient and integrated light vapor chip-scale platform, a few technological obstacles remain. First, the device should have long lifetime, i.e. very small rate of vacuum loss and very little outgassing. In addition, in order to enable scalability and reduced power consumption the volume of the active ACWGs should be minimized. Furthermore, one would like the encapsulation of the chip with a cell to be as compact as possible, i.e. a miniaturized vapor cell should be attached to chip. This could be achieved by anodic bonding a micro machined cell[29] to the surface of the chip. Additionally, for many applications, the device should have high optical density even while operating at low atomic density (i.e. non elevated temperatures). This will allow to reduce power consumption, increase abovementioned device longevity and reduce self broadening and shifts[30] accompanied with high density of the vapor. Finally, in order to fully harness the non-linear prospects (e.g. nonlinear all optical switching) of ACWGs, long ACWGs are desired. By using such long wave guides, one maintains the confined mode with the accompanied high energy density over long trajectories, and thus accumulate the non-linear effect. Our platform addresses some of the abovementioned requirements, as it offers an unprecedented small footprint (effective volume of $5 \cdot 10^{-4} mm^3$) enables operation with relatively low density of atoms and hence low power consumption and prolonged lifetime of chip based integrated vapor systems. Given the advanced work in the field, we are confident that pushing the boundaries of such technology by having a fully integrated light and vapor system on a chip will be accomplished in the near future.

**Acknowledgments**

The authors thank Avinoam Stern, Yefim Barash and Benny Levy from Accubeat Ltd. for the preparation of rubidium cells, and the use of the vacuum facilities and Prof. Tilman Pfau for fruitful discussions. The authors would like to acknowledge funding from the ERC grant LEVIN. The waveguides were fabricated at the Center for Nanoscience and Nanotechnology, The Hebrew University of Jerusalem

# Supplementary - Strong coupling and high contrast all optical modulation in atomic cladding waveguides

## 1. V-Type Pump-probe evanescent susceptibility

Here, we calculate the effective susceptibility of atomic vapor subject to a one dimensional evanescent pump and probe electromagnetic distribution. To do so, we follow the general derivation given in references 1 and 2, and modify it to account for an evanescent pump-probe configuration in the V-type configuration. Herby, we shall briefly describe the model, and its results in context of the main manuscript.

In figure S1a we describe the V-type level scheme. In the case presented in the accompanied manuscript, the |1> to |2> transition and |1> to |3> transition correspond to the 780nm D2 and 795nm D1 lines respectively. In figure S1b we describe schematically the scenario we are modeling. A one dimensional slab wave guide is illustrated as having a dielectric core, and an atomic vapor cladding. The electromagnetic mode that is supported by this wave guide, has a real wave number in direction x, and an imaginary wave number in the direction z. The later represents the evanescent wave. An atom is illustrated to have two distinct spatial-dynamic states: a) an atom with a *negative* velocity Vz (Vz is defined as positive in the direction of the z axis) entering the evanescent region. Such an atoms state will subsequently be described to be in steady state with a pump and probe that have an exponentially increasing electric field. Such a state is assumed to follow the spatial variation adiabatically, starting at the ground state far away from the surface. b) an atom with a *positive* Vz, which has left the surface at a quenched state, i.e, in it's ground state. Such an atom is subject to an abrupt change in its atomic state due to its collision with the surface (and consequently loss of coherence) in the presence of electromagnetic field, and thus the atom cannot be assumed to be in steady state.

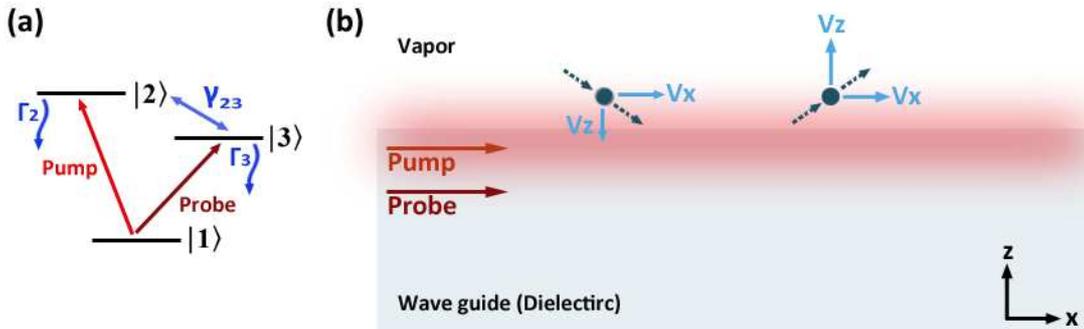

**Fig. S1:** a) V-type atomic level scheme b) Illustration of an atom entering the evanescent region of a one dimensional wave guide, with an negative velocity Vz, and an atom exiting the evanescent region with a positive velocity Vz.

It can be shown, that by separating the atoms into these two different groups one can write the following relation[1–3] for the effective susceptibility of the atomic cladding:

$$\chi = \frac{-2ik_z \hbar \Omega_s N}{I_s} \int_0^\infty dz \int dv W(v) e^{i2k_z z}[\Theta(v_z)\sigma_{13}(z,v) + \Theta(-v_z)\overline{\sigma_{13}}(z,v)] \quad (1)$$

Where $k_z$ is the imaginary z component wave number (k is the wave number in free space), $\Omega_s, I_s$ are the Rabi frequency and intensity of the probe, respectively, N is the atomic density, W is the Maxwell-Boltzmann velocity distribution, $\sigma_{13}(z,v)$ and $\overline{\sigma_{13}}(z,v)$ are the probes coherences for out coming and incoming atoms, respectively and $\Theta(v)$ is the Heaviside function. Thus, in order to calculate the susceptibility we are in need to find both $\sigma_{13}(z,v)$ and $\overline{\sigma_{13}}(z,v)$. We start by writing the general Bloch equations, in the spatial domain[2,4,5]:

$$\begin{cases} v_z \frac{d}{dz}\sigma_{11} = \frac{i\Omega_p(z)}{2}(\sigma_{21}-\sigma_{12}) + \frac{i\Omega_s(z)}{2}(\sigma_{31}-\sigma_{13}) + \Gamma_2\sigma_{22} + \Gamma_3\sigma_{33} \\ v_z \frac{d}{dz}\sigma_{22} = -\frac{i\Omega_p(z)}{2}(\sigma_{21}-\sigma_{12}) - \Gamma_2\sigma_{22} \\ v_z \frac{d}{dz}\sigma_{33} = -\frac{i\Omega_s(z)}{2}(\sigma_{31}-\sigma_{13}) - \Gamma_3\sigma_{33} \end{cases}$$

$$\left\{ v_z \frac{d}{dz}\sigma_{21} = \frac{i\Omega_p(z)}{2}(\sigma_{11}-\sigma_{22}) - \frac{i\Omega_s(z)}{2}\sigma_{23} + (i\Delta_p - \gamma_{21})\sigma_{21} \right.$$

$$\left\{ v_z \frac{d}{dz}\sigma_{31} = \frac{i\Omega_s(z)}{2}(\sigma_{11}-\sigma_{33}) - \frac{i\Omega_p(z)}{2}\sigma_{32} + (i\Delta_s - \gamma_{31})\sigma_{31} \right.$$

$$\left\{ v_z \frac{d}{dz}\sigma_{32} = \frac{i\Omega_s(z)}{2}\sigma_{12} - \frac{i\Omega_p(z)}{2}\sigma_{31} + (i(\Delta_s - \Delta_p) - \gamma_{32})\sigma_{32} \right.$$

Here $\Omega_s(z)$ and $\Omega_p(z)$ are the Rabi frequencies of the probe and pump respectively. Both are exponentially decaying along Z due to the evanescent electromagnetic field. $\sigma_{ij}$ is the slowly varying density matrix element. The detunings, and lifetimes are defined as follows:

$$\Gamma_i = \begin{cases} \Gamma_0 + |k_z v_z| & i \neq 1 \\ 0 & i = 1 \end{cases}$$

$$\gamma_{ij} = \frac{\Gamma_i + \Gamma_j}{2}$$

$$\Delta_s = \delta - k_x \cdot v_x \qquad \Delta_p = -k_x \cdot v_x$$

Where $\delta$ is the detuning of the probe beam from resonance, $k_z$ is the imaginary k-vector that describes the evanescent decay of the optical mode, and $k_x$ is the real propagation wave number. Here we assume that both modes, namely the pump and probes modes, have nearly the same wavenumbers. As we assume that $\Omega_p \gg \Omega_s$, and thus $\sigma_{11} + \sigma_{22} \approx 1$ (and thus $\sigma_{33} \approx 0$) we can write the following set of equations:

$$\left\{ v_z \frac{d}{dz}\sigma_{22} = -\frac{i\Omega_p}{2}(\sigma_{21}-\sigma_{12}) - \Gamma_2\sigma_{22} \right.$$

$$\begin{cases} v_z \frac{d}{dz}\sigma_{21} = \frac{i\Omega_p}{2}(1-2\sigma_{22}) + (i\Delta_p - \gamma_{21})\sigma_{21} \\ v_z \frac{d}{dz}\sigma_{12} = \frac{-i\Omega_p}{2}(1-2\sigma_{22}) + (-(i\Delta_p) - \gamma_{21})\sigma_{12} \end{cases}$$

$$\left\{ v_z \frac{d}{dz} \sigma_{31} = \frac{i\Omega_s}{2}(1 - \sigma_{22}) - \frac{i\Omega_p}{2}\sigma_{32} + (i\Delta_s - \gamma_{31})\sigma_{31} \right.$$

$$\left\{ v_z \frac{d}{dz} \sigma_{32} = \frac{i\Omega_s}{2}\sigma_{12} - \frac{i\Omega_p}{2}\sigma_{31} + (i(\Delta_s - \Delta_p) - \gamma_{32})\sigma_{32} \right.$$

To find such solutions, we solve $\sigma_{13}(z, v)$ numerically, given an exponential decaying $\Omega_s$ and $\Omega_p$, and an initial condition of $\sigma_{11} = 1$. For the steady state solution, we solve analytically the steady state (setting the derivatives to zero) set of equations.

In Fig. 2 we plot an example of such solutions, for the atoms moving along the z axis, i.e. those which satisfy $V_x=0$, and for a pump Rabi frequency of ~3GHz. We do so for two different velocities: in Fig 2a we plot the probe $\sigma_{31}$ coherence, as a function of distance from the surface for incoming and outgoing atoms with the speeds of -10m/s and +10m/s respectively (Fig 2a), and for atoms having incoming and outgoing speeds of -200m/s and 200m/s respectively (Fig 2b). As can be seen, such solutions differ both qualitatively, and quantitatively. For the slower atoms, the outgoing atoms experience a transient response, and undergo decaying Rabi oscillations, whereas the incoming atoms change their absorption in an adiabatic manner, up to a point where they start experiencing a strong pump beam (near the surface), and the induced transparency window starts to be evident. For the faster atoms, the outgoing atoms do not have sufficient time to develop Rabi oscillations, whereas the incoming atoms do not have sufficient time to be effected by the pump beam. Generally, in the case of a single photon experiment it has been pointed out [], that both atomic groups have an equal contribution to the susceptibility. In contrast, as has been pointed by ref. 2 , in our two photon scenario the two atomic groups do not contribute equally.

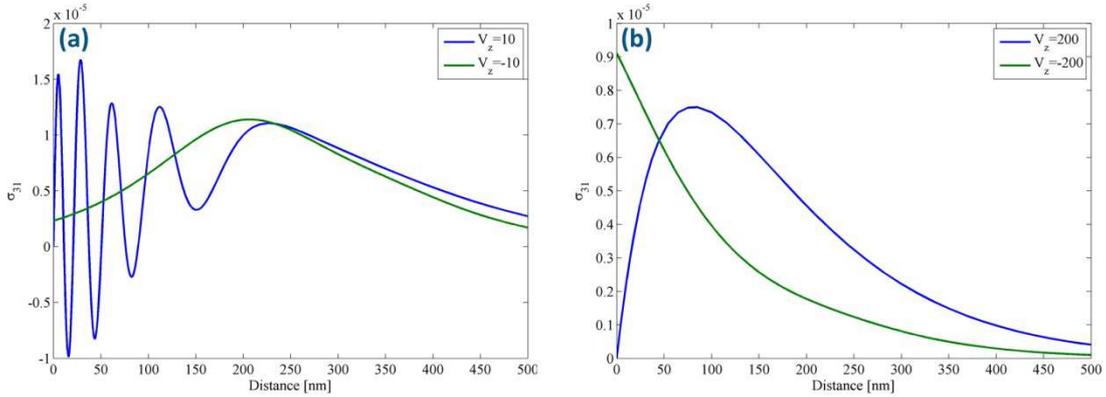

**Fig. S2**: a) $\sigma_{31}$ as function of distance from surface for the incoming group of atoms with speed of -10 m/s and outgoing speeds of 10m/s b) $\sigma_{31}$ as function of distance from surface for the incoming group of atoms with speed of -200 m/s and outgoing speeds of 200m/s

Finally, we plot the normalized susceptibility (given by Eq. 1). In Figure S3, we plot the solution for two cases. First, in figure S3a, we plot the solution for varying pump powers. As can be clearly seen the probe susceptibility is strongly influenced by the presence of the pump, and a transit time broadened transparency window is created in the center of susceptibility peak. Such results resemble the experimental results presented in the main article. Indeed we are able to nearly "shut down" the probe peak both in simulation and in the experiment. We note that the low pump power dips in the simulation are narrower than those measured. This is due to the fact that we have used a one-dimensional model, with a one dimensional transit time broadening, whereas our waveguide system is essentially two dimensional in respect to the transit time broadening. A two dimensional transit time broadening model shall broaden these lines significantly, as measured.

Finally, in order to verify that the strong dip within the susceptibility peak is indeed a coherent effect, we numerically "turn off" the coherent effects by nulling the $\sigma_{32}$ coherence matrix element and plot again the probe susceptibility under similar pump conditions (figure S3b). As can be clearly seen, now we are unable to reduce the probes susceptibility to below 50 percent of its maximal value. In general, in absence of $\sigma_{32}$ coherence, the only mechanisms to reduce of the probes susceptibility are atomic saturation and optical pumping. The absence of optical pumping discussed and justified in the main manuscript, and simulated here, leaves the saturation as the only mechanism for the reduction in the probe's susceptibility. As a consequence, in the absence of coherence, we are unable to achieve larger than 50 percent reduction in the susceptibility. As evident in the main text, the obtained contrast approaches 100% and thus coherent effects must contribute significantly. We note, that the small modulation dips that are evident at low pump powers in fig S3b are a consequence of a velocity selection saturation effect. This effect diminishes as power broadening increases.

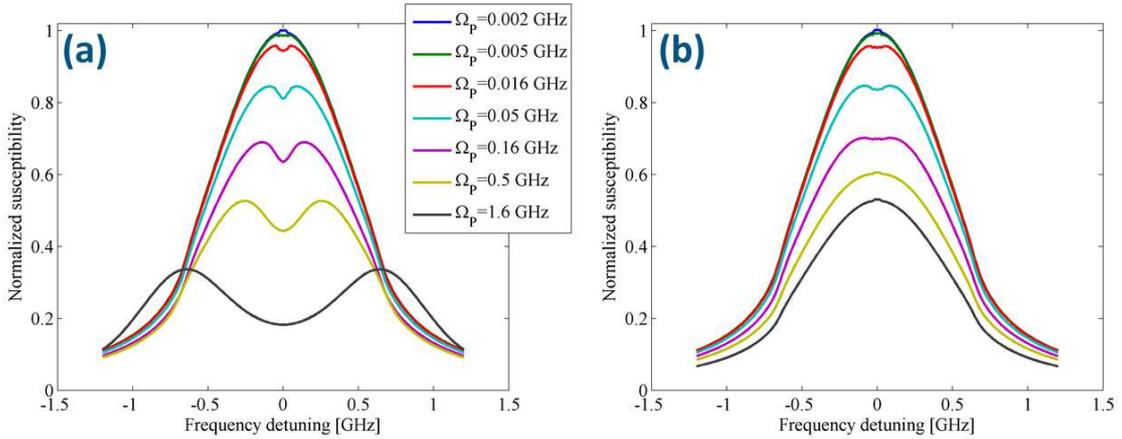

Fig. S3: a) Normalized atomic susceptibility as a function of frequency of evanescent V-type pump-probe excitation. The different curves correspond to varying pump powers. b) Normalized atomic susceptibility as function of frequency of evanescent V-type pump-probe excitation, where the coherence between levels 3 and 2 is set to zero.

## 2. Calculation of the Rabi frequencies

The purpose of this appendix is to present the calculations of the measured and theoretical Rabi frequencies presented in the main manuscript. First, we calculate the Rabi frequency as a function of the optical power in the wave guide.

Starting from the definition of Rabi frequency, and relating it to optical power:

$$|\Omega_r| = \left|\frac{d_{ij} \cdot E_0}{\hbar}\right| = \left(\frac{I}{2\epsilon_0 c}\right)^{\frac{1}{2}} \frac{d_{ij}}{\hbar} = \left(\frac{P/A}{2\epsilon_0 c}\right)^{\frac{1}{2}} \frac{d_{ij}}{\hbar}$$

Where, $d_{ij}$ is the dipole element, for a transition between levels $F_i$ and $F_j$, $\epsilon_0$ is the permittivity, c the speed of light, I is the optical intensity, P the optical power and A the area of the optical mode. Considering a mode area of 1200X250 nm², and taking the typical optical dipole element of the F=2 to F'=1/2/3 to be $1.1 \cdot 10^{-29} C \cdot m$ we obtain:

$$\frac{\Omega_r}{2\pi\sqrt{P}} = 419 \ [MHz/\sqrt{\mu W}]$$

Next, we relate the measured slope in Fig. 4b to a Rabi frequency, using the following relation:

$$\Delta f = \frac{1}{8\pi}\frac{\Omega_r^2}{\Delta\omega} = \frac{1}{8\pi}\frac{\alpha}{\Delta\omega}P$$

Where $\Delta f$ is the light shift, $\Omega_r^2 = \alpha P$ and $\Delta\omega$ is the detuning. Recalling that our measured slope was $13.49 \ MHz/\mu W$, we obtain

$$\alpha = 6.47 \cdot 10^6 \frac{MHz^2}{\mu W} \rightarrow \frac{\Omega_r}{2\pi\sqrt{P}} = \frac{\sqrt{\alpha}}{2\pi} = [405 \ MHz/\sqrt{\mu W}]$$